\documentclass[aps,showpacs,prx,amssymb,amsmath,twocolumn]{revtex4-2}
\usepackage{graphicx}
\usepackage{amssymb}
\usepackage{amsmath}
\usepackage{amsthm}
\usepackage{bm}
\usepackage{xcolor} 
\usepackage{array}
\usepackage{multirow}
\usepackage{tabularx}
\usepackage{booktabs}
\usepackage{textcomp}
\usepackage{mathtools}
\usepackage{hyperref}

\usepackage{qcircuit}

\usepackage{bbm}
\usepackage{cancel}
\usepackage{comment}
\usepackage{physics}

\begin{document}

\title{Quantum Computing Dataset of Maximum Independent Set Problem on King’s Lattice of over Hundred Rydberg Atoms}

\author{Kangheun Kim$^{1}$, Minhyuk Kim$^{1,2}$, Juyoung Park$^{1}$, Andrew Byun$^{1}$, Jaewook Ahn$^{1}$}
\email{jwahn@kaist.ac.kr}
\address {$^{1}$Department of Physics, KAIST, Daejeon 34141, Republic of Korea \\
$^{2}$Department of Physics, Korea University, Seoul 02841, Republic of Korea }

\begin{abstract}
Finding the maximum independent set (MIS) of a large-size graph is a nondeterministic polynomial-time (NP)-complete problem not efficiently solvable with classical computations. Here, we present a set of quantum adiabatic computing data of Rydberg-atom experiments performed to solve the MIS problem of up to 141 atoms randomly arranged on the King's lattice. A total of 582,916 events of Rydberg-atom measurements are collected for experimental MIS solutions of 733,853 different graphs. We provide the raw image data along with the entire binary determinations of the measured many-body ground states and the classified graph data, to offer bench-mark testing and advanced data-driven analyses for validation of the performance and system improvements of the Rydberg-atom approach.
\end{abstract}

\flushbottom
\maketitle

\thispagestyle{empty}

\section*{Background \& Summary}
The maximum independent set (MIS) problem belongs to the computational class of nondeterministic polynomial (NP)-complete problems, the hardest computational problems of no known classical algorithms that are efficient~\cite{erickson2019algorithms}. For a given graph $G(V,E)$, the MIS problem aims to find the maximum independent set, the largest set among the independent sets, where the independent set $I\subset V$ is defined as a set of unedged vertices, i.e., $v_i,v_j\in I$ and $(v_i,v_j)\notin E$. Rydberg quantum simulators are currently one of the biggest quantum computing physical platforms, capable of utilizing up to a few hundred qubits~\cite{ebadi2021quantum,scholl2021quantum}. In particular, the constraint of the independent set is implementable intrinsically with the Rydberg blockade effect that forbids two atoms proximate within a certain distance from being simultaneously excited to the same Rydberg-atom state~\cite{lucas2014ising,pichler2018quantum, urban2009observation,gaetan2009observation}. Therefore, a set of atoms arranged to a graph and simultaneously pumped to Rydberg atoms results in a non-adjacent arrangement of Rydberg atoms fulfilling the independent set constraint. In addition, by tuning the Hamiltonian of the atoms to maximize the number of Rydberg atoms, the MIS is achieved by the set of Rydberg atoms in the many-body ground state. In that regards, Rydberg-atom systems could perform adiabatic quantum computation (AQC) for the MIS problem in such a way that a target many-body ground state is prepared adiabatically from an easily preparable initial ground state. There are several recent experiments computing the solution of the MIS problem on the Rydberg-atom system by AQC~\cite{ebadi2022quantum,kim2022rydberg,byun2022finding}.

Here we provide a set of experimental AQC data of the MIS problem performed on the Rydberg-atom system. We first prepare an 11-by-18 array of optical tweezers. This lattice is identical to the union-jack-like king's graph~\cite{du1998handbook} recently experimented by Ebadi et al.,~\cite{ebadi2022quantum} in which its NP completeness on MIS problem has been addressed, {as in Fig.~\ref{fig:intro}. The MIS problem embedded on this graph has also been researched theoretically about its possibilities on quantum speedup~\cite{cain2023quantum,andrist2023hardness}. On 198 optical tweezer traps, atoms are stochasitcally loaded with about probability of about 50\%, and resulting random graphs are used. The size of experimented atom arrays is maximally 141 and on average about 104. Atoms then go through the initial state preparation stage, and the global laser field is applied to drive the total system for AQC. Laser intensity (Rabi frequency) and frequency (detuning) is slowly sweeped from the initial condition, and the initially prepared ground states are turned into the target final state before being measured with the imaging of the post AQC atomic array. There are 45 different experiment sets with different parameters including adiabatic sweep time, initial and final detunings, and for each set of data about 5,000 to 30,000 experiments were repeated. Besides the main data, we analyze major error sources from the measurement, Rydberg state decay, and the atom loss due to the finite temperature of the atoms. The pure effect of the control error scales with respect to the AQC sweeping time of an order of $\sim \tau^{-0.54(4)}$, comparable with the earlier reported $\sim\tau^{-0.48(2)}$ in Ebadi et al~\cite{ebadi2022quantum}. This data could be harnessed for the analysis of adiabatic computing behavior, the exploration of quantum phase transitions (QPT) in the transverse Ising model, and as a reference for conducting benchmark tests of the Rydberg atom approach to optimization problems.

\begin{figure*}[ht]
\centering
\includegraphics[width=0.8\linewidth]{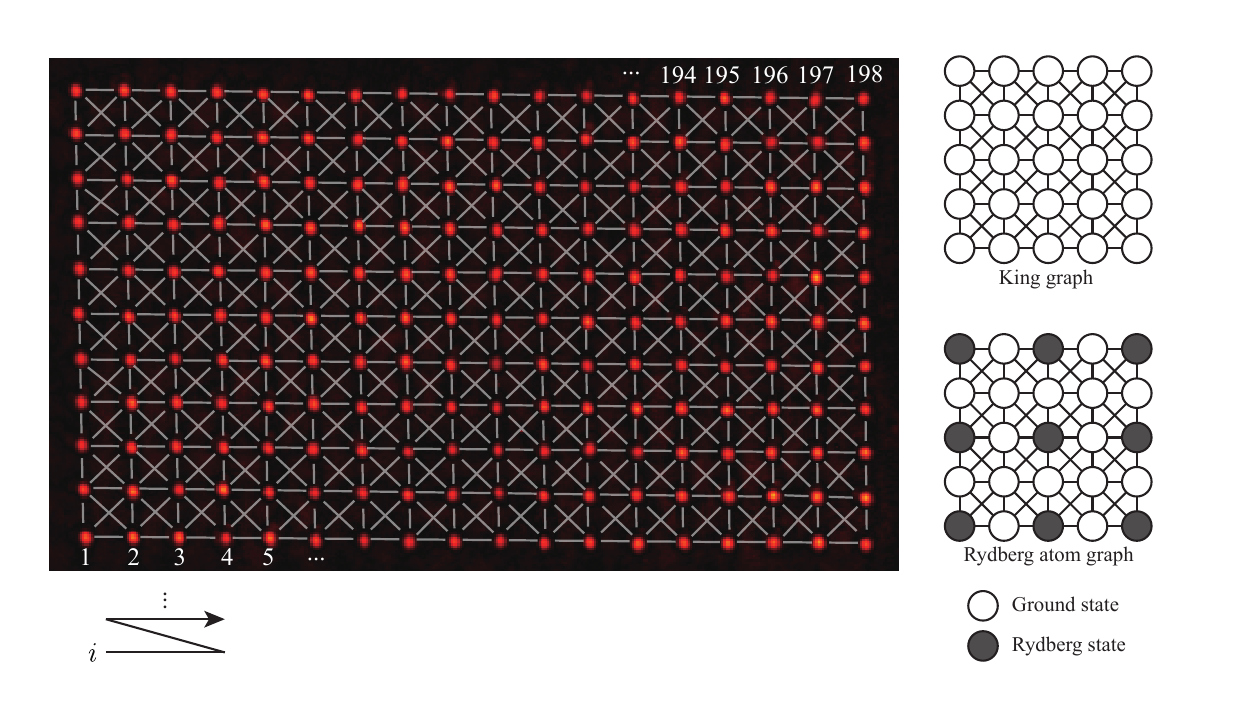}
\caption{Image of 198 optical tweezers with its index. Right side of the figure shows the connection between the atoms. The nearest atoms (6~$\mu$m) and the next nearest atoms (6$\sqrt{2}~\mu$m) are connected (Rydberg blockade radius $r_b\sim 10~\mu\rm m$), configuring king's graph.}
\label{fig:intro}
\end{figure*}

\section*{Methods}

\subsection*{Atom-array preparation } 
Rubidium atoms ($^{87}$Rb) are loaded on to an array of 18 by 11 optical tweezers, with the nearest atom distance of $d=6.0~\mu$m. Figure~\ref{fig:method1}(a) shows the experimental setup. To successfully trap the hundreds of atoms, all 198 tweezer traps are prepared as uniformly as possible. Tweezer traps are generated by Fourier transform of the phase displayed on the spatial light modulator (SLM, ODPDM512 by Meadowlark optics) shown in Fig~\ref{fig:method1}(a). The initial trap phase was generated with the Gerchberg-Saxton weighted (GSW) algorithm with 100 iterations starting from the uniform trap phase guess~\cite{kim2019gerchbergsaxton}. For a calibration of uniform traps, we measure the intensity of each 198 traps by (1) first image the  trap beam itself with the charged coupled device (CCD) (2) resonance scan the amount of a.c. Stark shift of the each trap site with the push-out beam ($5S_{1/2}, F=2 \rightarrow 5P_{3/2}, F'=3$). Meausred intensity is used to determine the weight given for each trap while iterating the GSW algorithm as $w_{i}=\langle I_i\rangle /(1-0.7(1-I_i/\langle I_i \rangle))$~\cite{nogrette2014singleatom}, where $I_i$ is the intensity of the $i$-th trap. We achieve the standard deviation below $3\%$ based on the atomic resonance. Initially inside the vacuum chamber, the $^{87}\rm Rb$ atoms are cooled and loaded inside the magneto optical trap (MOT) as an ensemble initially by turning on the anti-Helmholtz coil and the Doppler cooling beam. Optical tweezer beam from the Ti:Sapphire oscillator (TiC of Avesta) with wavelength 813 nm is turned on and atoms are loaded and by polarization gradient cooling (PGC) cooled down to $\sim 30~\mu\rm K$ inside the optical tweezer. We use objective lens with high numerical aperture (NA) of 0.5 (G Plan Apo 50X of Mitutoyo). Each trap has a beam radius of 1.1 $\mu$m, therefore each tweezer has about 1~mK depth in average. The trap depth data of all sites are recorded (see Data Records). The temperature is measured using the release and recapture method~\cite{tuchendler2008energy} before the main experiment. The temperature change due to laser conditions of the MOT beam causes different loss rate, which is considered in Technical Validation. Atoms are stochastically loaded in the tweezers with the per-site loading probability $\sim 0.5$, because of the collisional blockade~\cite{schlosser2002collisional}, resulting in a random geometry to be experimented for each repetition. Atom lifetime inside the trap is $\sim 40~$s and the experiment is operated with a repetition cycle of 1~s.

\begin{figure*}[ht]
\centering
\includegraphics[width=0.8\linewidth]{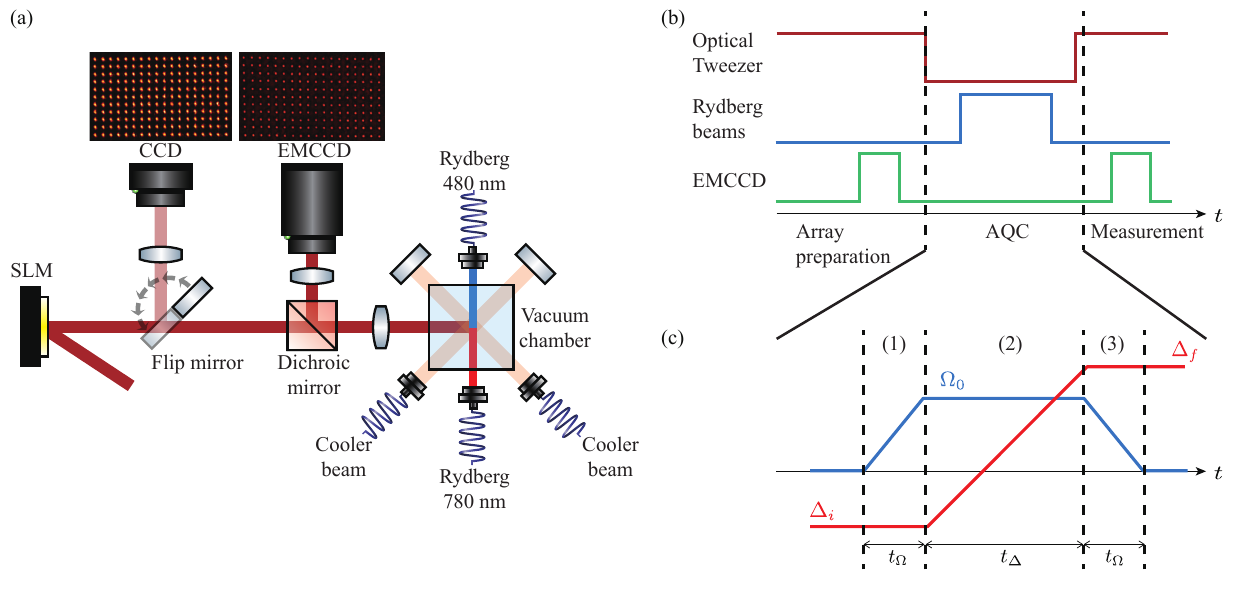}
\caption{(a) Experimental setup. (b) Experimental sequence. Loaded atoms are imaged initially, then the tweezer is turned off. The Rydberg operation is performed, and the tweezer is turned on, then results are imaged. (c) Adiabatic quantum computing sequence. Rabi frequencies are turned on and off linearly with time $t_{\Omega}$ with the maximum value $\Omega_0$, and the detuning is swept from $\Delta_i$ to $\Delta_f$ linearly during the time $t_\Delta$.}
\label{fig:method1}
\end{figure*}

\subsection*{AQC programming of MIS problem } Atoms are prepared in a random graph $G(V,E)$, in which a vertice in $V$ corresponds to each atom and an edge in $E$ corresponds to the van der Waals interaction between a strongly interacting atom pair. To be specific, we regard the atoms within the Rydberg blockade radius $r_b = (C_6/\Omega)^{1/6}\sim10~\mu \rm m$ are edged. The nearest neighbor distance is $6~\mu$m, and the next nearest neighbor distance (of diagonal edges) is $6\sqrt{2}\sim 8.5~\mu \rm m$, so the atoms are edged according to the union-jack-like king's graph as shown in Fig.~\ref{fig:intro}. Solving the MIS problem on the given graph is the same as finding the many-body ground state of the atom array in the MIS phase. The Hamitonian of the given Rydberg system consists is a transverse anti-ferromagnetic (AF) Ising Hamiltonian, given by
\begin{equation}
\hat{H}_{\rm Ryd} = \frac{\Omega(t)}{2} \sum_{i=1}^{N}{\hat{\sigma}^x_{i}} - \frac{\Delta(t)}{2} \sum_{i=1}^{N}{\hat{\sigma}^z_{i}} + \sum_{i,j}{U_{ij}\hat{n}_i\hat{n}_j},
\end{equation}
\noindent where $\hslash = 1$, $\Omega$ is Rabi frequency, $\Delta$ is detuning, $U_{ij} = C_6/R_{ij}^{6}$ is is the distance $R_{ij}$ dependent interaction for $C_6 = 2\pi\times 1023~{\rm GHZ/}\mu{\rm m}^{-6}$, and $\hat{n}_i=(\hat{\sigma}^z_i+1)/2$ defined for a psedo-spinor system of $\ket{g}=\ket{5S_{1/2},F=2,m_F=2}$ and $\ket{r} = \ket{71S_{1/2},m_J=1/2}$. The hamiltonian is driven and controlled with the two-photon transition of 780 nm and 480 nm beams. For $\Omega = 0$, $\Delta <0$, the many-body ground state is the paramagnetic $\ket{g}^{\otimes N}$. When $\Omega = 0$, $0<\Delta<U_{diag}$, the many-body ground state is in the MIS phase~\cite{byun2022finding}.

The experimental sequence diagram, which includes AQC, is depicted in Fig.~\ref{fig:method1}(b). AQC takes place when the optical tweezers are turned off, and photographs are taken of each randomly configured graph along with the final results both before and after the AQC process. We program our AQC to sweep the atom array from the paramagnetic $\ket{g}^{\otimes N}$ phase to the MIS phase. Initially, we prepare the initial ground state $\ket{g}^{\otimes N}$ by optical pumping. Then we take three passages, as in Fig.~\ref{fig:method1}(c), changing $(\Omega,\Delta)$ from (1) $(0,\Delta_i)\rightarrow (\Omega_0,\Delta_i)$ for $t_\Omega$, (2) $(\Omega_0,\Delta_i) \rightarrow (\Omega_0,\Delta_f)$ for $t_\Delta$, and (3) $(\Omega_0,\Delta_f) \rightarrow (0,\Delta_f)$ for $t_\Omega$ while turning off the trap. For our experiment, we fix $t_\Omega = 0.3~\mu \rm s$ and $\Omega_0$ is position dependent along the y-axis of the lattice because the beam is Gaussian, and the values of $\Omega_0$ for different positions are calculated from Rabi freuqency measurements as in Table.~\ref{tab:Rab} (see Technical Validations). We then turn off the trap for $6.5~\mu \rm s$ for an AQC experiment. There are 45 distinct experiments characterized by varying $t_\Delta$, $\Delta_i$, and $\Delta_f$, each having data of 4,000 to 30,000 repetitions. The experiments are summarized in Table.~\ref{tab:exp}. For the all 45 experiments, the final detuning $\Delta_f$ is larger than $0$ and smaller than the diagonal interaction $U_{\rm diag} = C_6/(8.5~\mu{\rm m})^6 = 2\pi \times 2.7 ~\rm MHz$, satisfying the MIS conditions.

\begin{table}[ht]
\caption{\label{tab:Rab} Average y pixel and Rabi frequency for each of 11 rows.}
\centering
\begin{ruledtabular}
\begin{tabular}{ccc}
Row \# & $\langle y_{\rm pixel}\rangle $ & $\langle \Omega \rangle_{\rm Row \#}/2\pi$ (MHz) \\
\hline
1& 47 & 0.961 \\
2& 67 & 1.09 \\
3& 86 & 1.23 \\
4& 106 & 1.34 \\
5& 125 & 1.41 \\
6& 145 & 1.41 \\
7& 164 & 1.35 \\
8& 185 & 1.23 \\
9& 204 & 1.09 \\
10& 223 & 0.962 \\
11& 244 & 0.856 \\
\end{tabular}
\end{ruledtabular}
\end{table}

\subsection*{Identification of random atom array and AQC measurement. } To obtain the information of the initial atom array and the AQC result, two images of the atom array embedded in the tweezers Fig.~\ref{fig:method3}(a) are captured in each experimental iteration using the electron-multiplied charged coupled device (EMCCD, iXon Ultra 888 by Andor). The first photo, e.g., in Fig~\ref{fig:method3}(b), is taken following array preparation, where atoms are loaded randomly and the second photo, e.g., in Fig~\ref{fig:method3}(c), is taken after the AQC process. During each image taking, the fluorescence of each trap site is analyzed to determine whether an atom is trapped inside the optical tweezer. Discriminating between the photon counts of trapped and non-trapped atoms provides digitized data for the experiment, as in  Fig~\ref{fig:method3}(d). Notably, for the second photo taken after AQC, the trap is activated before the photograph, causing atoms in the Rydberg state to be anti-trapped from the optical tweezers. Therefore, the remaining atoms in the second photo are considered to be in the ground state $\ket{g}$, while the disappeared atoms are regarded as being in the Rydberg state $\ket{r}$.

\begin{figure*}[ht]
\centering
\includegraphics[width=0.8\linewidth]{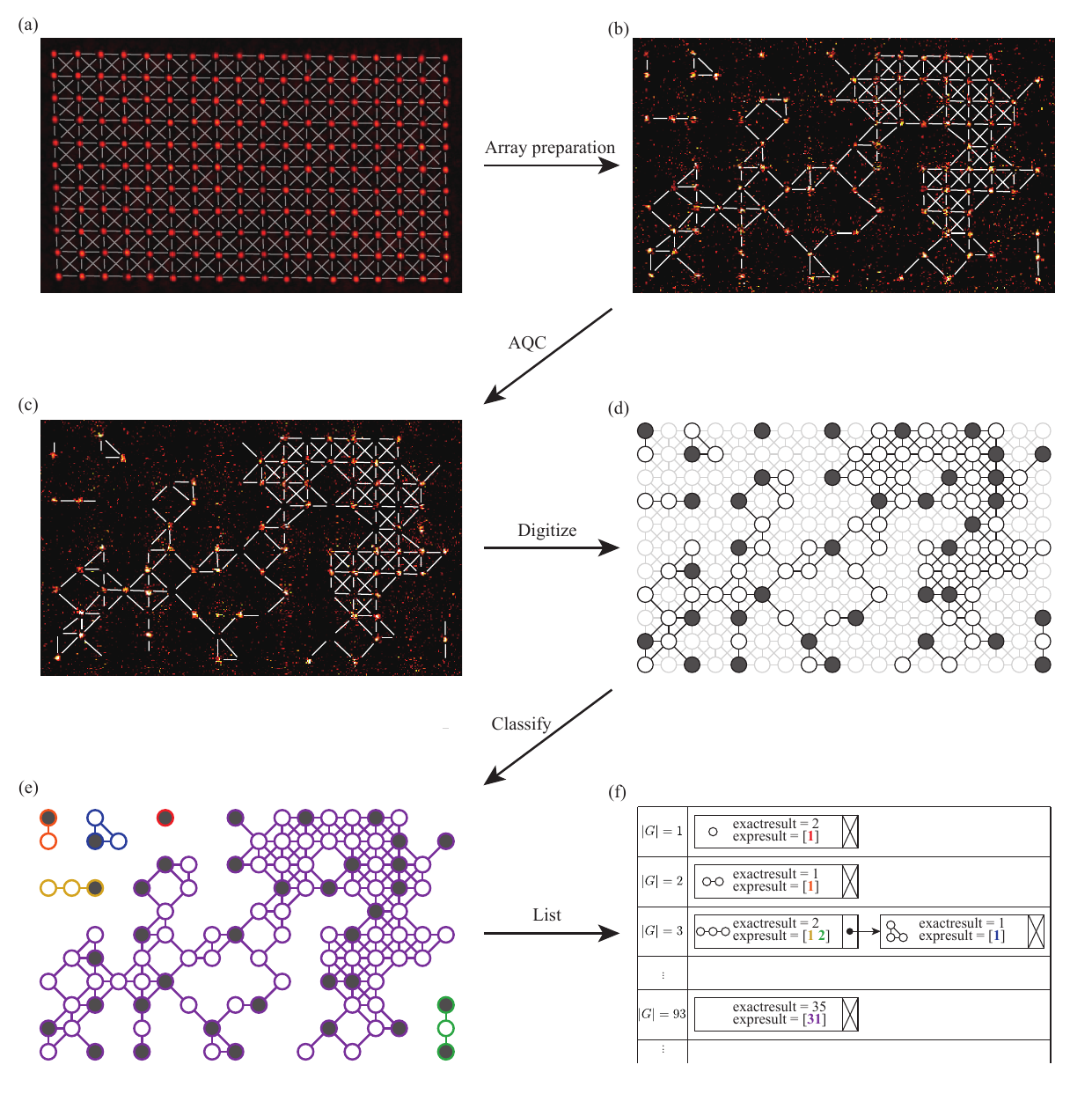}
\caption{Data processing of quantum computing data (a) Initial 198 optical tweezer sites (b) Atoms are randomly loaded into tweezer with a probability of $\sim 0.5$ (c) Atoms after the AQC. Rydberg atoms are anti-trapped. (d) Digitize the prepared array (circles) and the Rydberg atoms (filled circles) with the images from (b) and (c). (e) Classify the graph into connected components (colored if connected) (f) List all connected graphs, same sizes linked by the linked list structure, where each node contains a graph, exact MIS solution, experimented MIS solutions (by the list), and the graph hash.}
\label{fig:method3}
\end{figure*}

\section*{Data Records}

The compilation of 45 distinct experimental parameters can be found in Table.~\ref{tab:exp}. The complete dataset is accessible through the following {\it figshare} link: \hyperlink{https://doi.org/10.25452/figshare.plus.c.6938829}{https://doi.org/10.25452/figshare.plus.c.6938829}. {For the seperate links;} Raw image data \hyperlink{https://doi.org/10.25452/figshare.plus.24503680}{https://doi.org/10.25452/figshare.plus.24503680} for experiments 1 to 30, and \hyperlink{https://doi.org/10.25452/figshare.plus.24523198}{https://doi.org/10.25452/figshare.plus.24523198} for experiments 31 to 45. Additionally, the raw images for release-and-recapture measurements and Rabi frequency calibration are available at \hyperlink{https://doi.org/10.25452/figshare.plus.24540988}{https://doi.org/10.25452/figshare.plus.24540988}. Fluorescence data, trap position data, and trap depth data can be accessed through \hyperlink{https://doi.org/10.6084/m9.figshare.23828004}{https://doi.org/10.6084/m9.figshare.23828004}. The GraphTable data, which is explained below, is provided at \hyperlink{https://doi.org/10.6084/m9.figshare.23911413}{https://doi.org/10.6084/m9.figshare.23911413}.

\begin{table}[ht]
\caption{\label{tab:exp}Initial, final detunings, detuning sweep time, $p_{rg}$ number, and repetition number of the experiment.}
\centering
\begin{ruledtabular}
\begin{tabular}{cccccc}
Exp 	&$\Delta_i/2\pi$ 	&$ \Delta_f/2\pi $	&$ t_{\Delta}$ 	&$p_{rg,\rm SPAM}$ & Rep \\
\# 		& (MHz)			&  (MHz)			&  ($\mu$s)		& \# 					&  \# \\
\hline
1&-4.0 & 2.0 & 0.1 & 3 & 7,000 \\
2&-4.0 & 2.0 & 0.2 & 3 & 7,000 \\
3&-4.0 & 2.0 & 0.3 & 3 &7,000 \\
4&-4.0 & 2.0 & 0.4 & 3 & 7,000 \\
5&-4.0 & 2.0 & 0.5 & 1 & 32,999\\
6&-4.0 & 2.0 & 0.6 & 3 & 11,000 \\
7&-4.0 & 2.0 & 0.7 & 3 & 11,000 \\
8&-4.0 & 2.0 & 0.8 & 3 & 11,000 \\
9&-4.0 & 2.0 & 0.9 & 3 & 8,248 \\
10&-4.0 & 2.0 & 1.0 & 1 & 30,000 \\
11&-4.0 & 2.0 & 1.5 & 1 & 26,543 \\
12&-4.0 & 2.0 & 2.0 & 1 & 29,321 \\
13&-4.0 & 2.0 & 2.5 & 1 & 27,558 \\
14&-4.0 & 2.0 & 3.0 & 1 & 23,788 \\
15&-4.0 & 2.0 & 3.5 & 1 & 22,476 \\
16&-4.0 & 2.0 & 4.0 & 1 & 23,797 \\
17&-4.0 & 2.0 & 4.5 & 1 & 19,742 \\
18&-4.0 & 2.0 & 5.0 & 1 & 25,392 \\
19&-6.0 & 2.0 & 1.0 & 3 & 4,946 \\
20&-5.5 & 2.0 & 1.0 & 3 & 3,618 \\
21&-5.0 & 2.0 & 1.0 & 3 & 4,000 \\
22&-4.5 & 2.0 & 1.0& 3& 4,000 \\
23&-4.0 & 2.0 & 1.0& 3 & 4,000 \\
24&-3.5 & 2.0 & 1.0& 3 & 4,000 \\
25&-3.0 & 2.0 & 1.0& 3 & 4,000 \\
26&-4.0 & 2.5 & 0.5 & 2 & 14,154 \\
27&-4.0 & 2.5 & 1.0 & 2 & 13,763 \\
28&-4.0 & 2.5 & 1.5 & 2 & 12,948 \\
29&-4.0 & 2.5 & 2.0 & 2 & 11,283 \\
30&-4.0 & 2.5 & 2.5 & 2 & 8,898 \\
31&-4.0 & 2.5 & 3.0 & 2 & 9,431 \\
32&-4.0 & 2.5 & 3.5 & 2 & 9,991 \\
33&-4.0 & 2.5 & 4.0 & 2 & 5,636 \\
34&-4.0 & 2.5 & 4.5 & 2 & 3,479 \\
35&-4.0 & 2.5 & 5.0 & 2 & 8,510 \\
36&-4.4 & 1.1 & 0.5 & 3 &14,675\\
37&-4.4 & 1.1 & 1.0 & 3 & 13,522\\
38&-4.4 & 1.1 & 1.5 & 3 &15,000 \\
39&-4.4 & 1.1& 2.0 & 3 &11,350\\
40&-4.4 & 1.1& 2.5 & 3 &10,000\\
41&-4.4 & 1.1& 3.0 & 3 &10,000\\
42&-4.4 & 1.1& 3.5 & 3 &7,411\\
43&-4.4 & 1.1 & 4.0 & 3 &13,438 \\
44&-4.4 & 1.1& 4.5 & 3 &15,000 \\
45&-4.4 & 1.1& 5.0 & 3 &15,000 \\
\end{tabular}
\end{ruledtabular}
\end{table}

The primary data is provided in the form of `.tif' and `.mat' files, including raw image files, fluorescence data for each atom, and digitized data for each atom. Raw image files are in TIF format, constituting a list of image files structured as a three-dimensional array. The first two dimensions denote the image, while the third dimension represents the stack of images across experimental repetitions. Due to the substantial file size, raw data is segmented into multiple files for the same experiments, named as `Exp\{\textit{Exp\#\}}\_X\{\textit{\#}\}.tif', where \textit{Exp\#} corresponds to the experiment index in Table~\ref{tab:exp}, and \textit{\#} is assigned in sequential order. Fluorescence data and digitized data are stored in MATLAB data files, featuring the three-dimensional array variables `fluoAreshape' and `fluodigreshape,' alongside the variable `threshold,' which establishes the threshold for fluorescence data digitization following the outlined methods. The 3D array data in `flouAreshape' and `floudigreshape' is organized as follows: the first dimension denotes each tweezer, the second dimension signifies the two photographs of each repetition, and the third dimension corresponds to the number of experiment repetitions. For instance, in MATLAB language, extracting the 2D array of the experiment for the first repetition from `fluodigreshape' is achieved by using `squeeze(fluodigreshape(:,:,1))'. The first column of this array presents digitized data from the first photo, while the second column represents data from the second photo. A value of `1' indicates that the atom is trapped, while a value of `0' signifies that the atom is not trapped. Consequently, in the second column, a `1' implies measurement in the ground state $\ket{g}$, whereas a `0' signifies measurement in the Rydberg state $\ket{r}$.

The data for the $\ket{g}\rightarrow\ket{r}$ SPAM error is labeled as `SPAM\{\textit{$p_{rg,\rm SPAM}$\#}\}', encompassing both the raw image file in `.tif' format and the data file in `.mat' format. The image file contains experiments related to release and recapture data, while the MATLAB data file incorporates variables such as `fluoAreshape', `fluodigreshape', along with two one-dimensional arrays, namely `Prg' and `counts'. Within the MATLAB data, `Prg' captures the measurement error of $\ket{g}\rightarrow \ket{r}$ for each atom, as determined by the release and recapture method (see Technical Validation). The `counts' array stores the counts associated with the measurement of each `Prg' data. It is noted that \textit{$p_{rg,\rm SPAM}$\#} serves as the index for $p_{rg,\rm SPAM}$ data, varying across experiments due to distinct cooling laser conditions (see Technical Validation). The trap depth data is provided in a 1D MATLAB file named `trap\_depth.mat', featuring an array with the same name. Each value in this array corresponds to the a.c. Stark shift of each trap site. For instance, the initial value, $17.6691$, signifies that the first trap site has an a.c. Stark shift of $2\pi\times 17.6691 ~\rm MHz$. The raw data for Rabi frequency calibration (see Technical Validation) is contained in the image file `RabiFrequencyCalibration.tif', while its processed data can be found in the MATLAB file `RabiFrequencyCalibration.mat'.

We organize both the exact solution and the experimentally obtained solutions for each randomly configured (isomorphic) graph. Initially, the graphs are categorized into connected components, as illustrated in Fig.\ref{fig:method3}(e). Subsequently, for each connected graph, the exact solutions are determined using the `MISSolver.py' code (see Code Availability). The resulting data is structured in a table implemented as a linked list class named `GraphTable', and stored within the Python script `GraphTable.py' (Fig.\ref{fig:method3}(f), see Code Availability). The instances of the generated `GraphTable' are then saved by the \textit{pickle} module of Python under the file name of `Exp{\textit{Exp\#}}.pkl'. Each `GraphTable' instance represents a Python list of the linked list class `GraphLinkedList' (saved in `GraphLinkedList.py', see Code Availability), arranged according to the graph size $|G|$. Within each `GraphLinkedList', there exists a linked list pertaining to the same graph size $|G|$. Each instance of the node class (`GraphNode') encapsulates the data of each isomorphic graph, comprising the graph structure, exact solutions, and a Python list detailing the experimented solutions.

\section*{Technical Validation}
In Fig.~\ref{fig:techval}(a), the histogram illustrates log events organized by both the graph size $G$ and the corresponding errors $\delta_G$ for each graph $G$ in Experiment 10 as an example. The error $\delta_G$ for a given graph $G$ is calculated as the difference between the exact MIS solution and the experimental MIS solution (Rydberg excitation number). The major error sources in the AQC include the state-preparation-and-measurement (SPAM) error and the control error. In the technical validation section, we assess the SPAM error's magnitude and establish the validity of our data by analyzing the behavior of the control error while varying the sweep speed and final detuning of the AQC. SPAM errors are single-site bit-flip errors following a binomial distribution. Consequently, their effect on the MIS solution of the graph $G(V,E)$ is, on average, proportional to the graph size $|G|$ itself. In the quasi-adiabatic regime, the control error adheres to the quantum Kibble-Zurek mechanism (QKZM)~\cite{damski2005simplest,zurek2005dynamics,damski2006adiabaticimpulse,keesling2019quantum,ebadi2021quantum}, which is also proportional to the graph size $|G|$. Thus, the overall error in the graph $G(V,E)$ can be modeled on average as ${\langle}\delta_{G}\textcolor{red}{\rangle} = (\alpha p_{gr}-(1-\alpha)p_{rg})|G|$, where $\alpha$ represents the average solution ratio, and $p_{gr}$ and $p_{rg}$ denote the error rates for transitions from $\ket{r}\rightarrow \ket{g}$ and $\ket{g}\rightarrow \ket{r}$, respectively.

\begin{figure*}[htb]
\centering
\includegraphics[width=0.8\linewidth]{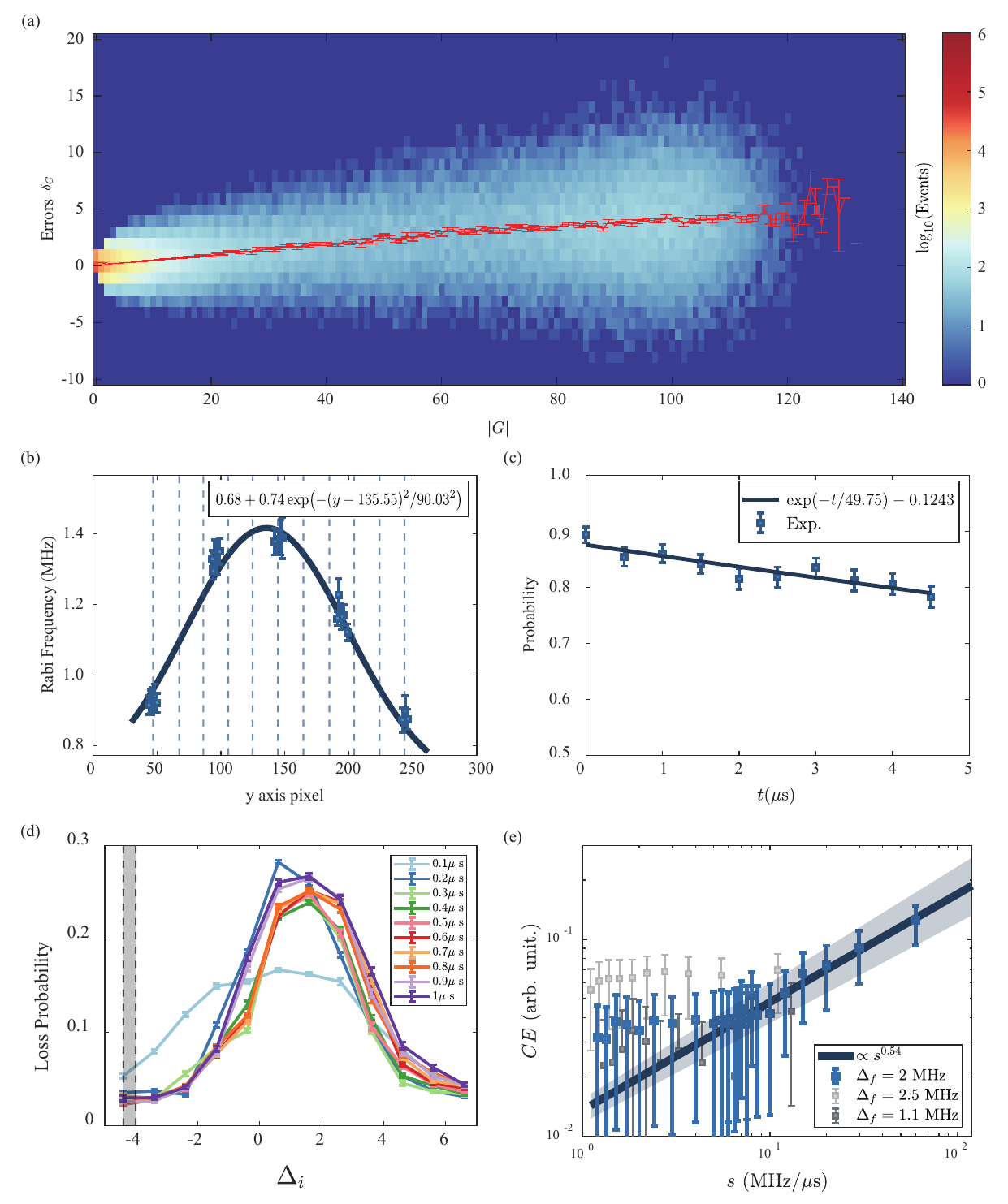}
\caption{Figures for technical validation (a) Plot of log events with regards to the graph size $|G|$ and the error $\delta_G$ for `Exp10'. Here the error $\delta_G$ is the difference between the exact MIS solution and the experimented MIS solution. (b) Rabi frequency for each $y$ pixel. Average values for each row are given in Table~\ref{tab:Rab}. (c) Measurement of Rydberg state decay time. Rydberg probability after $\pi$ pulse and a time $t$ is measured. (d) Calibration for initial detuning $\Delta_i$. From the sequence of Fig.~\ref{fig:method3}(c), the sequence (2) is deleted, and $\Delta_f=\Delta_i$ for the calibration experiment, is expected to have no loss. Shaded region of $\Delta_i$ and $t_\Omega = 0.3~\mu\rm s$ are used for our main experiment, where loss probability is small. (e) Control error parameter $CE$ as a function of sweep rate $s$. Fitting for $s>7.5~{\rm MHz}/\mu\rm s$ gives $\mu = 0.54(4)$ comparable with the value from other group $0.48(2)$~\cite{ebadi2022quantum}.}
\label{fig:techval}
\end{figure*}

\subsection*{Rabi frequency for each atoms } 
Due to the finite radius of the Gaussian Rydberg beam, the Rabi frequencies of individual atoms exhibit variations. To see the difference of Rabi frequencies among atoms, we measure the Rabi frequency on 25 atoms, each positioned sufficiently far apart to ensure independence (with a minimum distance of $17~\mu$m, where $r_B = 10~\mu$m). The Rabi frequency for each $y$ axis position is determined through fitting based on this data, allowing us to calculate the Rabi frequency for the 198 atoms in the main experiment as in Fig~\ref{fig:techval}(b). Additionally, we provide the average calculated Rabi frequency for each of the 18 rows of the experiment in Table~\ref{tab:Rab} for reference. In the subsequent discussion on control error, we do not account for the differences in control error resulting from varying Rabi frequencies but instead focus on the average control error.

\subsection*{SPAM errors } SPAM errors originating from $\ket{r}\rightarrow \ket{g}$ ($p_{gr,\rm SPAM}$) and $\ket{g}\rightarrow \ket{r}$ ($p_{rg,\rm SPAM}$) stem from distinct sources. The SPAM error from $\ket{r}\rightarrow \ket{g}$ ($p_{gr,\rm SPAM}$) is predominantly attributed to the decay of the Rydberg state. Conversely, the error from $\ket{g}\rightarrow \ket{r}$ ($p_{rg,\rm SPAM}$) primarily results from atom loss in the trap, stemming from factors such as the finite temperature of atoms and non-resonant scattering, among others~\cite{deleseleuc2018analysis}. To characterize these errors, we initially examine the overlap of the fluorescence data at each atom site. Subsequently, we measure the Rydberg decay time, and finally, we directly measure the loss rate by releasing the atom and subsequently recapturing it.

For the measurement of the Rydberg decay time $T_1$, we initiate a $\pi$ pulse, followed by a wait period of $t$, and then apply another $\pi$ pulse. By varying the duration $t$, we observe the decay of the measurement, as depicted in Fig.\ref{fig:techval}(c). Fitting the data with the model $e^{-t/T_1}+c$, we obtain $T_1 = 49(18)~\mu$s. While ideal theoretical considerations, accounting for black body radiation, suggest a lifetime of approximately $150 \mu \rm s$\cite{beterov2009quasiclassical}, realistic errors have led to similar values, with other groups reporting $T_1=50(8)\mu \rm s$\cite{levine2018highfidelity}. For our experiment, where the trap is turned off for $6.5~\mu \rm s$, we can calculate $p_{gr,\rm SPAM} = 1-e^{-6.5/49(18)} = 0.12^{+0.06}_{-0.03}$.

To directly measure the atom loss rate stemming from factors like finite temperature and non-resonant laser scattering, the release and recapture method is employed~\cite{tuchendler2008energy}. After trapping the atom, we turn the trap off for a duration of $6.5~\mu\rm s$, aligning with the trap turning-off time in the main experiment. Subsequently, the atom is recaptured by reactivating the trap laser. The recapture rate is measured, and the loss rate $p_{rg,\rm SPAM}$ is determined as 1 minus the recapture rate. Three distinct $p_{rg,\rm SPAM}$ measurements ($p_{rg, \rm SPAM}$ \# in Table~\ref{tab:exp}) are presented, reflecting variations in cooling laser power conditions leading to different Polarization Gradient Cooling (PGC) rates and, consequently, distinct loss rates. In each instance, the mean loss rates are 0.0730, 0.0877, and 0.0966, respectively. The data for these three different loss rate measurements are also provided for each atom under the name `SPAM\textit{\#}.mat' (see Data Records). To characterize the potential nonuniformity in the total optical tweezer array, which could result in a nonuniform atom loss rate for each trap site, the trap depth is measured for each tweezer site. The trap depth data is available in the `trapdepth.mat' file, presented in frequency units (MHz).

\subsection*{Control errors }

\subsubsection*{Early stage control error measurement for $t_{\Omega}$ and $\Delta_i$. } 
To validate that our data exhibits minimal errors during the initial sweep of the measurement, we conducted experiments as depicted in Fig.~\ref{fig:techval}(d). During this process, the laser is both turned on (ramping $\Omega$ from $\Omega_0$ to 0) and turned off (ramping $\Omega$ from 0 to $\Omega_0$) while maintaining a fixed detuning at the initial value $\Delta_i$. In the absence of control errors during this stage, there should be no atom loss as the system returns to its initial condition. We varied the detuning from $-4.4\rm MHz$ to $6.6~\rm MHz$ with increments of 1~MHz and adjusted the ramping time $t_{\Omega}$ from $0.1$~$\mu$s to $1~\mu$s with a step size of $0.1~\mu$s. The loss probability is calculated as one minus the ratio of remaining atoms compared to the initially captured atoms. For the main experiment, we selected $t_{\Omega} = 0.3~\mu \rm s$ and $\Delta_i = -4.4~\rm MHz$ and $-4~\rm MHz$, as these parameters yielded the lowest loss probabilities among the considered parameters. Therefore, we can affirm that our main experiment is robust to early-stage control errors.

\subsubsection*{Control error estimates from the main experiment data }
Control errors are investigated by conducting experiments with varying sweep rates in the sequence. The control errors in our experiment align with the universal scaling predicted by the Quantum Kibble-Zurek Mechanism (QKZM) in relation to the sweep speed. The sweep speed $s$ is defined as $s=(\Delta_f -\Delta_i)/t_{\Delta}$, where $t_\Delta$ is the detuning sweep time. Experiments 1 to 25, 26 to 35, and 36 to 45 can be grouped into experimental sets to examine control errors by altering specific parameters ($s$ or $\Delta_f$) while keeping others constant. The average total error is modeled as previously, $\textcolor{red}{\langle}\delta_{G}\textcolor{red}{\rangle} = (\alpha p_{gr}-(1-\alpha)p_{rg})|G|$. This linear behavior can be observed in the initial segment of Fig.~\ref{fig:techval}(a) (red line). We can now express the equation as follows:
\begin{eqnarray}
\label{eq3}
CE &\equiv& \alpha p_{gr,\rm control}-(1-\alpha) p_{rg,\rm control} \nonumber \\
& =&  \frac{\textcolor{red}{\langle}\delta_{G}\textcolor{red}{\rangle}}{|G|} +\left((1-\alpha) p_{rg,\rm SPAM}- \alpha p_{gr,\rm SPAM}\right)
\end{eqnarray}
We introduce the control error parameter $CE$ to represent the second term of the equation, which characterizes the control error value in the experiment. The value $CE$ can be computed using the last term of the equation, which is accessible through experimental data. The average solution rate, denoted as $\alpha$, can be calculated from the exact solutions for each graph obtained in the experiment, with a fixed value of $\alpha\simeq 0.43$ regardless of the sweep speed. The value $\textcolor{red}{\langle}\delta_G\textcolor{red}{\rangle}/|G|$ represents the slope of the red line in Fig~\ref{fig:techval}(a) and varies with the sweep speed. Fig.\ref{fig:techval}(e) illustrates the plot of $CE$ for all experiments, where various sweep rates $s$ were explored. In this context, it is clear that when $\Delta_f$ is held constant, the control error in our data follows the anticipated pattern: the control error diminishes for smaller sweep speeds $s$ or extended annealing times $s^{-1}$, while it escalates for larger sweep speeds $s$ or abbreviated annealing times $s^{-1}$. Furthermore, we can fit the control error as a function of the sweep rate $s$ due to QKZM, yielding $CE\propto s^{\mu}$ with $\mu =0.54(4)$ for the fitting range $s=7.5{\rm MHz}/\mu \rm s$ to $60~{\rm MHz}/\mu \rm s$. This value is in line with earlier research, which reported $\mu=0.48(2)$ using the same lattice (supplementary of Ref.~\cite{ebadi2022quantum}). We note that the slope values $\textcolor{red}{\langle}\delta_G\textcolor{red}{\rangle}/|G|$ and the ratio $\alpha$ are obtained by using $|G|<50$ data.

\section*{Usage Notes}
The dataset is accessible using MATLAB. Python users can access the data using the \textit{scipy.io.loadmat} function within the \textit{scipy} library. Analysis of randomly generated graphs can be performed using the `networkx' library in Python. The data can be grouped from 1 to 18, 19 to 25, 26 to 35, and 36 to 45, considering that only one parameter differs within each group.

\section*{Code availability}
The raw data can be utilized to generate fluorescence data (`flouAreshape') and digitized data (`floudigreshape') using the provided code (`Digitze\_Data.m') available on \textit{figshare} (\hyperlink{https://doi.org/10.6084/m9.figshare.23911368}{https://doi.org/10.6084/m9.figshare.23911368}). The data processing involved in Technical Validations entails the classification of randomly generated graphs by verifying group isomorphism and obtaining the true solution for the MIS problem. This task is accomplished using custom Python code, also provided. The MIS solver is included in the `MISSolver' class, saved in the Python script `MISSolver.py'. This solver is based on the \textit{github} code~\cite{liu2019misalgorithms} and its corresponding reference~\cite{fomin2009measure}. The `GraphTable' class, found in Data Records, is stored in the Python script `GraphTable.py'. Instances of `GraphTable' are saved using the Python \textit{pickle} module. The `GraphLinkedList' class and a `GraphNode' class are preserved in the Python script `GraphLinkedList.py'. The code for saving the experimented solution to the `GraphTable' instances is implemented in the Python Jupyter notebook `Save\_Graphtable.pynb'. An example code for reading the `.pkl' instance is provided in `Open\_Graphtable\_Example.pynb'.

\bibliography{MD_reference}

%

\end{document}